# Appearance of bulk Superconductivity under Hydrostatic Pressure in $Sr_{0.5}RE_{0.5}FBiS_2$ (RE=Ce, Nd, Pr and Sm) compounds

Rajveer Jha, Brajesh Tiwari and V.P.S. Awana[*]

CSIR-National Physical Laboratory, Dr. K.S. Krishnan Marg, New Delhi-110012, India

**Abstract**

We report the appearance of superconductivity under hydrostatic pressure (0–2.5GPa) in $Sr_{0.5}RE_{0.5}FBiS_2$ with RE=Ce, Nd, Pr and Sm. The studied compounds, being synthesized by solid state reaction route, are crystallized in tetragonal *P4/nmm* space group. At ambient pressure, though the RE=Ce exhibits the onset of superconductivity below 2.7K, the Nd, Pr and Sm samples are not superconducting down to 2K. With application of hydrostatic pressure (up to 2.5GPa), superconducting transition temperature ($T_c$) is increased to around 10K for all the studied samples. The magneto-resistivity measurements are carried out on all the samples under 2.5GPa pressure and their upper critical fields ($H_{c2}$) are determined. The superconductivity of these compounds appears to be quite robust against magnetic field. Summarily, the $Sr_{0.5}RE_{0.5}FBiS_2$ compounds with RE=Ce, Nd, Pr and Sm are successfully synthesized and superconductivity is induced in them under hydrostatic pressure.

[*]**Corresponding Author**
Dr. V. P. S. Awana, Principal Scientist
E-mail: awana@mail.nplindia.org
Ph. +91-11-45609357, Fax-+91-11-45609310
Homepage awanavps.wenbs.com

**Introduction**

BiS$_2$ based superconductivity was first observed at around 4.5K in BiS$_2$ layers of Bi$_4$O$_4$S$_3$ compound [1,2]. Later the same was seen in REO/FBiS$_2$ [3-12] compounds. In short span of time, the BiS$_2$ based superconductivity attracted tremendous interest from the scientific community [1-12]. The principal reason for the same is that the layered structure of these compounds is exactly similar to that of famous REO$_{1-x}$F$_x$FeAs pnictide high $T_c$ superconductors. In case of REO$_{1-x}$F$_x$BiS$_2$ the charge carriers are introduced by O$^{2-}$ site F$^{1-}$ substitution in REO redox layer and superconductivity is established in adjacent BiS$_2$ layer. This is exactly same as being done in case of REO$_{1-x}$F$_x$FeAs, where O/F substitution in REO redox layer introduces superconductivity in adjacent FeAs [13]. It is also reported, that the superconductivity of BiS$_2$ based REO$_{1-x}$F$_x$BiS$_2$ compounds enhances tremendously under hydrostatic pressure [14-17]. The increase in superconducting transition temperature ($T_c$) has been reported even up to five fold or so. For example the $T_c$ of LaO$_{0.5}$F$_{0.5}$BiS$_2$ increases from around 2.1K to above 10K under hydrostatic pressure of say above 2GPa [14,18]. Researchers also synthesized the high pressure high temperature (*HPHT*) processed compounds of REO$_{1-x}$F$_x$BiS$_2$ series and established stable superconductivity of up to 10K [16]. As far as theoretical aspects are concerned, detailed electronic structure calculations are done and the role of sp orbitals of Bi and S is highlighted [19-21]. The situation seems to be in parallel with Fe and As hybridization in case of Fe pnictide (REO$_{1-x}$F$_x$FeAs) superconductors [13]. No doubt, the BiS$_2$ based recent superconductivity is catching the attention of scientific community [1-12, 14-22].

It is proposed earlier that SrFBiS$_2$ is iso-structural to LaOBiS$_2$, which may become superconducting with suitable doping [23]. In SrFBiS$_2$, the carriers are introduced by Sr$^{2+}$ site La$^{3+}$ substitution, and superconductivity is achieved of up to 2.5K [24-26]. Also, the evolution of superconductivity is reported by substitution of tetravalent Th$^{+4}$, Hf$^{+4}$, Zr$^{+4}$, and Ti$^{+4}$ for trivalent La$^{+3}$ in LaOBiS$_2$ compound [27]. Interestingly, though in case of REO$_{1-x}$F$_x$BiS$_2$ superconductivity is reported for RE=La, Pr, Nd and Ce [3-12], in SrFBiS$_2$ series only Sr$_{1-x}$La$_x$BiS$_2$ is yet studied [24-26]. It was thus required to check upon the SrFBiS$_2$ series with Sr site different RE substitutions. Recently we found that superconducting transition temperature ($T_c$) of Sr$_{0.5}$La$_{0.5}$FBiS$_2$ compound increases by five-fold from around 2K to above 10K, accompanied with a semiconducting to metallic transformation in normal state under just above 1GPa external pressure [28]. Keeping all this in mind, in current article, we report successful synthesis of $Sr_{0.5}RE_{0.5}FBiS_2$ with RE=Ce, Nd, Pr and Sm, and study their superconducting properties with and without application of hydrostatic pressure. It is found that though RE=Nd, Pr and Sm do not show superconductivity down to 2K, the RE=Ce exhibits the onset of superconductivity below 2.5K. Interestingly, all the studied compounds demonstrated superconductivity of up to 10K under hydrostatic pressure. Their magneto-transport measurements showed that these are quite robust against magnetic field. Here we report on the synthesis and appearance of bulk superconductivity in $Sr_{0.5}RE_{0.5}FBiS_2$ systems with different RE (RE=Ce, Nd, Pr and Sm) other than La.

**Experimental details**

The bulk polycrystalline samples of series $Sr_{0.5}RE_{0.5}FBiS_2$ (RE=Ce, Nd, Pr and Sm) were synthesized by solid state reaction route via vacuum encapsulation. High purity Ce, Nd, Pr, Sm, $SrF_2$, Bi and S were weighed in stoichiometric ratio and ground in pure Argon atmosphere filled glove box. The mixed powders were subsequently palletized and vacuum-sealed ($10^{-4}$ mbar) in quartz tubes. The box furnace was used to sinter the samples at $650^oC$ for 12h with the typical heating rate of $2^oC/min$. The sintered samples were subsequently cooled down slowly to room temperature. This process was repeated twice. X-ray diffraction (*XRD*) patterns were recorded for all samples at room temperature in the scattering angular (*2θ*) range of $10^o$-$80^o$ in equal *2θ* step of $0.02^o$ using *Rigaku Diffractometer* with $CuK_\alpha$ (*λ*=1.54Å) radiation. Rietveld analysis was performed for all samples using the standard *FullProf* program.

The pressure dependent resistivity measurements were performed on Physical Property Measurements System (*PPMS*-14Tesla, *Quantum Design*) using *HPC*-33 Piston type pressure cell with *Quantum design DC* resistivity Option. Hydrostatic pressures were generated by a BeCu/NiCrAl clamped piston-cylinder cell. The sample was immersed in a pressure transmitting medium (Daphne Oil) covered with a Teflon cell. Annealed Pt wires were affixed to gold-sputtered contact surfaces on each sample with silver epoxy in a standard four-wire configuration.

**Results**

The room temperature observed and Reitveld fitted *XRD* patterns for the $Sr_{0.5}RE_{0.5}FBiS_2$ (RE=Ce, Nd, Pr and Sm) compounds are shown in figure 1. All the compounds are crystallized in tetragonal structure with space group *P4/nmm*. Small impurity peak of $Bi_2S_3$ is also observed in $Sr_{0.5}Sm_{0.5}FBiS_2$ compound, which is close to the background of *XRD* pattern. On the other hand all the other studied samples are mostly phase pure within the *XRD* limit. Rietveld fitted lattice parameters are $a=4.065(1)Å$, $c=13.35(2)Å$ for $Sr_{0.5}Ce_{0.5}FBiS_2$; $a=4.056(3)Å$, $c=13.38(2)Å$ for $Sr_{0.5}Nd_{0.5}FBiS_2$; $a=4.061(2)Å$, $c=13.36(1)Å$ for $Sr_{0.5}Pr_{0.5}FBiS_2$ and $a=4.053(1)Å$, $c=13.40(3)Å$ for $Sr_{0.5}Sm_{0.5}FBiS_2$ compounds. It is worth mentioning that for the Sm-based compound the intensity (profile) of peak around 26.5° i.e., reflection (004) does not fit along with some other reflections like (005) and (007). These observations indicate a preferred orientation along *c*-crystallographic axis and/or impact of Sm atomic form factor, which may not be true for other RE compounds. We stick to the nominal composition for the refinement as changing the occupancy did not improve the fitting. The volume of the unit cell obtained through the Rietveld fitting analysis of each compound and other parameters, including their atomic co-ordinate positions etc. are summarized in the table 1. The volume of unit cell is increasing with the increasing ionic radii of the RE element from Sm to Ce. There is a possibility of internal chemical pressure on the unit cell of $Sr_{0.5}RE_{0.5}FBiS_2$, similar to that as in the case of $REO_{0.5}F_{0.5}BiS_2$ [3,5,10] with different RE.

The temperature dependent electrical resistivity $\rho(T)$ for the $Sr_{0.5}RE_{0.5}FBiS_2$ (RE=Ce, Nd, Pr and Sm) compounds in the temperature range 2-300K is shown in figure 2. All the compounds show semiconducting behavior down to 2K, except the $Sr_{0.5}Ce_{0.5}FBiS_2$, which exhibits the superconducting transition ($T_c^{onset}$) near 2.7K, while $\rho=0$ is not attained down to 2K, see figure 2. It is expected that the superconducting transition temperature ($T_c$) may increase with replacing La to Nd, Ce, Pr and Sm, as in the case of $REO_{0.5}F_{0.5}BiS_2$ [3,10]. The $T_c^{onset}$ is absent down to 2K in these compounds, except for RE=Ce, which exhibits $T_c^{onset}$ near 2.7K. Interestingly, superconducting transition temperature ($T_c$) for similar structure Fe pnictides REFeAsO/F [13,29] and $BiS_2$ based $REO/FBiS_2$ systems [3-5,8-10] scales from around 26K for La to 55K for Sm in former (Fe-pnictides) and 2.1K (La) to above 5K (Nd) in later ($BiS_2$ based ones). The obvious RE ionic size dependence of $T_c$ is not visible in currently studied $Sr_{0.5}RE_{0.5}FBiS_2$ system, and the reason behind the same is not clear. Seemingly, the chemical pressure on the unit cell is not large enough to bring in superconductivity in these compounds at ambient pressure. This is the reason that we applied external pressure on the studied $Sr_{0.5}RE_{0.5}FBiS_2$ compounds and achieved superconductivity in them, which will be discussed in figure 4(a-d).

Figure 3(a-d) depicts the temperature dependence of *DC* magnetic susceptibility (field cool) from 300K down to 2K in applied field of 1kOe for $Sr_{0.5}RE_{0.5}FBiS_2$ (RE=Ce, Nd, Pr and Sm). Curie-Weiss law, $\chi=C/(T-\Theta)$, fitting was performed to determine the paramagnetic effective magnetic moments ($\mu_{eff}$) and dominant magnetic ordering (sign of $\Theta$). The estimated effective magnetic moments are; $\mu_{Ce}$=2.97$\mu_B$/f.u. (C=1.19emu.K/mol-Oe), $\mu_{Nd}$=2.87$\mu_B$/f.u. (C=1.429emu.K/mol-Oe), $\mu_{Pr}$=3.14$\mu_B$/f.u. (C=1.24emu.K/mol-Oe), $\mu_{Sm}$=3.24$\mu_B$/f.u. (C=1.31emu.K/mol-Oe) for $Sr_{0.5}Ce_{0.5}FBiS_2$, $Sr_{0.5}Nd_{0.5}FBiS_2$, $Sr_{0.5}Pr_{0.5}FBiS_2$ and $Sr_{0.5}Sm_{0.5}FBiS_2$ respectively and their corresponding Curie-Weiss temperature ($\Theta$) are -216.31K, -72.61K, -59.18K and -160.90K. The negative $\Theta$ for all the compounds suggests that the predominant magnetic ordering is antiferromagtic. A clear magnetic transition can be seen in the upper inset of the figures 3(a) and (d) for $Sr_{0.5}Ce_{0.5}FBiS_2$ and $Sr_{0.5}Sm_{0.5}FBiS_2$ at around 8K, which does not saturates down to 2K under 1kOe. After this submission in a preprint [30] for $Sr_{0.5}Ce_{0.5}FBiS_2$, a similar transition is observed at around 8K, which was claimed to be of ferromagnetic nature being coexisting with superconductivity. Since the obtained Curie-Weiss temperature is negative, the predominant ordering seems to be antiferromagnetic, which can be further complimented with isothermal magnetization (*M-H*) curves at 2K. The *M-H* plots are shown in the lower insets of figure 3(a-d), which do not indicate any sign of saturation up to 5kOe, though small loop opening is seen for RE=Ce and Sm, suggesting the canting of magnetic moments in these samples. A clear paramagnetic behavior is observed for $Sr_{0.5}Nd_{0.5}FBiS_2$ and $Sr_{0.5}Pr_{0.5}FBiS_2$ from the magnetization curves at 2K, as can be viewed in the insets of figure 3(b) and (c) respectively.

Figure 4(a-d) shows temperature dependent electrical resistivity $\rho(T)$ at various applied pressures of 0-2.5GPa in the temperature range 2-300K for the $Sr_{0.5}RE_{0.5}FBiS_2$ (RE=Ce, Nd, Pr and Sm) compounds. Though normal state resistivity shows the semiconducting behavior up to applied hydrostatic pressure of 1.5GPa for all the samples, but the resistivity values are suppressed remarkably. Interestingly, the normal state resistivity changes from semiconducting to the metallic one at 2.0 and 2.5GPa applied pressures for all the samples. Inset of the figure

4(a-d) i.e., the magnified view of resistivity curves show clear superconducting transitions at various pressures. Apart from $Sr_{0.5}Ce_{0.5}FBiS_2$, the other compounds did not show superconducting onset at ambient pressure. Sufficient broadening in the superconducting transitions is observed under applied pressure of 1.5GPa for $Sr_{0.5}Ce_{0.5}FBiS_2$, $Sr_{0.5}Nd_{0.5}FBiS_2$ and $Sr_{0.5}Sm_{0.5}FBiS_2$ and at 0.97GPa for $Sr_{0.5}Pr_{0.5}FBiS_2$. Also the onset of superconducting transition ($T_c^{onset}$) increases with increasing pressure.

Figure 5 shows the pressure dependence of the superconducting transition temperature ($T_c^{onset}$) for the studied $Sr_{0.5}RE_{0.5}FBiS_2$ (RE=Ce, Nd, Pr and Sm) compounds. It can be inferred, that superconductivity gradually increases with pressure up to the 0.97GPa, for all the samples. For higher pressure of 1.5GPa, the $T_c^{onset}$ shoots up to above 9K. It seems that 1.5GPa is the transition pressure ($P_t$) for all the samples. With further increase in the pressure, though the $T_c^{onset}$ gets almost saturated, the $T_c(\rho=0)$ is increased, resulting in relatively sharper superconducting transition widths in the pressure range of 1.5-2.5GPa. In our results we also observed slight decreases in $T_c^{onset}$ for $Sr_{0.5}Ce_{0.5}FBiS_2$ and $Sr_{0.5}Sm_{0.5}FBiS_2$ samples at the 2.5GPa pressure; otherwise the trend of enhancement of superconductivity under pressure is same for all the samples.

Figure 6(a-d) shows the $\rho(T)$ under applied magnetic fields (1-7Tesla) for the $Sr_{0.5}RE_{0.5}FBiS_2$ (RE=Ce, Nd, Pr and Sm) under applied pressure of 2.5GPa. It is worth noting that with increasing magnetic field, the $T_c(\rho=0)$ decreases more rapidly as compared to the $T_c^{onset}$. In these compounds the superconducting transition width is broadened under applied magnetic fields, which is similar to high-$T_c$ cuprates and Fe-based pnictide superconductors. We summarize the magneto-resistivity results for all studied compounds as critical fields verses temperature phase diagrams, which is being shown in figure 7(a-d). The critical fields are estimated at corresponding temperatures, where resistivity drops to 90%, 50% and 10% of the normal state resistance [$\rho_n(T,H)$] in applied magnetic fields. The upper critical field, $H_{c2}(T)$ (using 90% criteria) at absolute zero temperature $H_{c2}(0)$ is determined by the conventional one-band Werthamer–Helfand–Hohenberg (*WHH*) equation, i.e., $H_{c2}(0)=-0.693T_c(dH_{c2}/dT)_{T=Tc}$. The estimated $H_{c2}(0)$ are 13.8Tesla, 11.6Tesla, 13.5Tesla and 14.5Tesla for $Sr_{0.5}Ce_{0.5}FBiS_2$, $Sr_{0.5}Nd_{0.5}FBiS_2$, $Sr_{0.5}Pr_{0.5}FBiS_2$ and $Sr_{0.5}Sm_{0.5}FBiS_2$ respectively. In figure 7 (a-d), the solid lines are the extrapolation to the Ginzburg–Landau equation $H_{c2}(T)=H_{c2}(0)(1-t^2/1+t^2)$, where t=$T/T_c$ is the reduced temperature, which gives values slightly higher than that by *WHH* approach. These upper critical field values for all samples are close to but within the Pauli paramagnetic limit i.e., $H_p=1.84T_c$. The liquid vortex region between irreversible field $H_{irr}(T)$ and upper critical field $H_{c2}(T)$ is of significant importance for a superconductor. $H_{irr}$ is determined using 10% criteria of magnetoresistivity and is slightly less than half of the upper critical field.

**Discussion**

Rare-earth substituted BiS$_2$-based layered compounds $Sr_{0.5}RE_{0.5}FBiS_2$ (RE=Ce, Nd, Pr and Sm) are crystallized in tetragonal structure with space group *P4/nmm*. The choice of different RE-ions in these $Sr_{0.5}RE_{0.5}FBiS_2$ (RE=Ce, Nd, Pr and Sm) compounds is based on the fact that superconductivity has yet been reported only for $Sr_{0.5}La_{0.5}FBiS_2$ [24-26], which curiously increased from 2.5K to 10K under external hydrostatic pressure [28]. It is important to note, that there is nearly no effect of different RE-ions on the *c* lattice parameter however *a* decreases with the ionic radii of RE-ion in $Sr_{0.5}RE_{0.5}FBiS_2$. Under ambient pressure the studied compounds show semiconductor like conduction in normal state with minimum resistivity for $Sr_{0.5}Pr_{0.5}FBiS_2$ at room temperature. Apart from $Sr_{0.5}Ce_{0.5}FBiS_2$, none of the other studied compounds showed onset of superconducting transition down to 2K. Under hydrostatic pressure starting from 0.25GPa to 2.50GPa, all these compounds showed superconductivity with $T_c^{onset}$ reaching up to 9-10K. The sharp increase in $T_c^{onset}$ under up to 1.5GPa pressure and its further saturation for higher pressures suggest possible structural change in these Pauli paramagnetic limited ($H_p\sim1.84T_c$) superconductors with applied hydrostatic pressure. Based on structural and magnetic studies, Tomita et.al [31] have argued in similar BiS$_2$-based compound (LaO$_{1-x}$F$_x$BiS$_2$), that a sudden increase of $T_c$ with pressure is related to the structural transition from tetragonal to monoclinic. The significant increases in the $T_c$ as well as the suppression of semiconducting behavior in the normal state resistivity suggest increase in the charge carrier density in the pressure range of 0-1.5GPa. In similar structure, LnO$_{0.5}$F$_{0.5}$BiS$_2$ (Ln-La, Ce) compounds, C. T. Wolowiec et. al., have suggested the possibility of coexistence of two superconducting phases near 1.0GPa pressure [16]. In a study on single crystal of PrO$_{0.5}$F$_{0.5}$BiS$_2$ the normal state resistivity changed from semiconducting-like behavior to metallic behavior under pressure and $T_c$ increased from 3.5K to 8.5K [32]. A recently appeared preprint based on detailed first principle calculations by Morice et.al, for a similar set of compounds i.e., LnO$_{1-x}$F$_x$BiS$_2$ (Ln=La,

Ce, Pr, and Nd), especially the CeOBiS$_2$ concluded that these compounds may display a pressure induced semiconducting to metal transition and also the change of rare-earth does not affect the Fermi surface [33]. There is a possibility of enhancement of strong electron correlations as well within same crystallographic phase under pressure. The pressure dependent structural studies are warranted on presently studied Sr$_{0.5}$RE$_{0.5}$FBiS$_2$ (RE=Ce, Nd, Pr and Sm) compounds. This could help in knowing the possible reason behind dramatic changes in their transport properties, including appearance of superconductivity and normal state semiconducting to metallic transformation under moderate pressure of around 1.5GPa. The resistivity under magnetic field at the pressure of 2.5GPa exhibited a gradual decrease in $T_c$ with increasing magnetic field, suggesting that these compounds are type-II superconductors. The upper critical fields being determined by *WHH* model are close to but within the Pauli paramagnetic limit i.e. $H_p=1.84T_c$. Also the upper critical field is almost double to as compared to irreversibility field at absolute zero temperature. The verification of these observations for layered fluorosulfide Sr$_{0.5}$RE$_{0.5}$FBiS$_2$ (RE=Ce, Nd, Pr and Sm) may need further magnetometry studies under hydrostatic pressure.

In summary, we have synthesized layered fluorosulfide compounds Sr$_{0.5}$RE$_{0.5}$FBiS$_2$ (RE=Ce, Nd, Pr and Sm), which are crystallized in tetragonal *P4/nmm* space group. These compounds are semiconducting in the temperature range 300-2K at ambient pressure and become metallic under pressures above 1.5GPa. All the compounds show superconductivity under hydrostatic pressure with $T_c^{onset}$ of 10K. In resistivity under magnetic field measurements at the applied pressure of 2.5GPa, these compounds appear to be quite robust against magnetic field.

**Acknowledgements**

Authors would like to thank their Director NPL India for his keen interest in the present work. This work is financially supported by *DAE-SRC* outstanding investigator award scheme on search for new superconductors. Rajveer Jha acknowledges the *CSIR* for the senior research fellowship.

Table 1

Atomic coordinates volume of unit cell, and Wyckoff positions, for studied $Sr_{0.5}RE_{0.5}FBiS_2$ compounds

| Samples → | $Sr_{0.5}Ce_{0.5}FBiS_2$ | $Sr_{0.5}Nd_{0.5}FBiS_2$ | $Sr_{0.5}Pr_{0.5}FBiS_2$ | $Sr_{0.5}Sm_{0.5}FBiS_2$ |
|---|---|---|---|---|
| $\chi^2$ | 4.19 | 3.54 | 4.58 | 6.33 |
| *Bragg R-factor* | 4.96 | 2.64 | 3.68 | 6.20 |
| *Rf-factor* | 3.52 | 1.82 | 3.43 | 4.68 |
| $a=b$(Å) | 4.065(1) | 4.056(3) | 4.061(2) | 4.053(1) |
| $c$(Å) | 13.35(2) | 13.38(2) | 13.36(1) | 13.40(3) |
| $c/a$ | 3.28 | 3.29 | 3.28 | 3.30 |
| $V$(Å$^3$) | 221.11(2) | 220.22(3) | 220.52(1) | 220.35(3) |
| Sr/RE(1/4, 1/4, z) | 0.110(1) | 0.109(2) | 0.114(1) | 0.113(1) |
| Bi(1/4, 1/4, z) | 0.623(3) | 0.624(2) | 0.624(3) | 0.626(1) |
| S1(1/4, 1/4, z) | 0.379(1) | 0.381(1) | 0.364(2) | 0.363(3) |
| S2(1/4, 1/4, z) | 0.809(2) | 0.812(3) | 0.810(1) | 0.822(1) |
| F(x,y,z) | (3/4,1/4,0) | (3/4,1/4,0) | (3/4,1/4,0) | (3/4,1/4,0) |

**Figure Captions**

**Figure 1:** (Color online) Observed (*open circles*) and calculated (*solid lines*) XRD patterns of $Sr_{0.5}RE_{0.5}FBiS_2$ (RE=Ce, Nd, Pr and Sm) compound at room temperature.

**Figure 2:** (Color online) Resistivity versus temperature $\rho(T)$ plots for $Sr_{0.5}RE_{0.5}FBiS_2$ compounds, at ambient pressure in the temperature range 300K-2K. Inset shows the $\rho(T)$ curve in the temperature range 6-2K.

**Figure 3:** (a-d) (Color online) Temperature dependent magnetic susceptibility $\chi(T)$ for (a) $Sr_{0.5}Ce_{0.5}FBiS_2$, (b) $Sr_{0.5}Nd_{0.5}FBiS_2$, (c) $Sr_{0.5}Pr_{0.5}FBiS_2$ and (d) $Sr_{0.5}Sm_{0.5}FBiS_2$ compounds at ambient pressure. The upper inset of (a and d) is zoomed part in the temperature range 2-10K and lower inset of each (a-d) shows the isothermal magnetization at 2K.

**Figure 4:** (a-d) (Color online) Resistivity versus temperature $\rho(T)$ plots at various applied pressures (0-2.5GPa) in the temperature range 300K-2.0K, inset of each shows the zoomed part of $\rho(T)$ plots near the superconducting transition temperature for (a) $Sr_{0.5}Ce_{0.5}FBiS_2$, (b) $Sr_{0.5}Nd_{0.5}FBiS_2$, (c) $Sr_{0.5}Pr_{0.5}FBiS_2$ and (d) $Sr_{0.5}Sm_{0.5}FBiS_2$ compounds.

**Figure 5:** $T_c^{onset}$ vs applied pressure for the $Sr_{0.5}RE_{0.5}FBiS_2$ compounds.

**Figure 6:** (a-d) (Color online) Temperature dependence of the Resistivity $\rho(T)$ under magnetic fields at 2.5GPa hydrostatic pressure for (a) $Sr_{0.5}Ce_{0.5}FBiS_2$, (b) $Sr_{0.5}Nd_{0.5}FBiS_2$, (c) $Sr_{0.5}Pr_{0.5}FBiS_2$ and (d) $Sr_{0.5}Sm_{0.5}FBiS_2$ compounds.
.

**Figure 7:** (a-d) (Color online) The temperature dependence of upper critical field $H_{c2}$ and $H_{iir}$, determined from 90% and 10% resistivity criterion respectively for (a) $Sr_{0.5}Ce_{0.5}FBiS_2$, (b) $Sr_{0.5}Nd_{0.5}FBiS_2$, (c) $Sr_{0.5}Pr_{0.5}FBiS_2$ and (d) $Sr_{0.5}Sm_{0.5}FBiS_2$ compounds.

**Fig. 1**

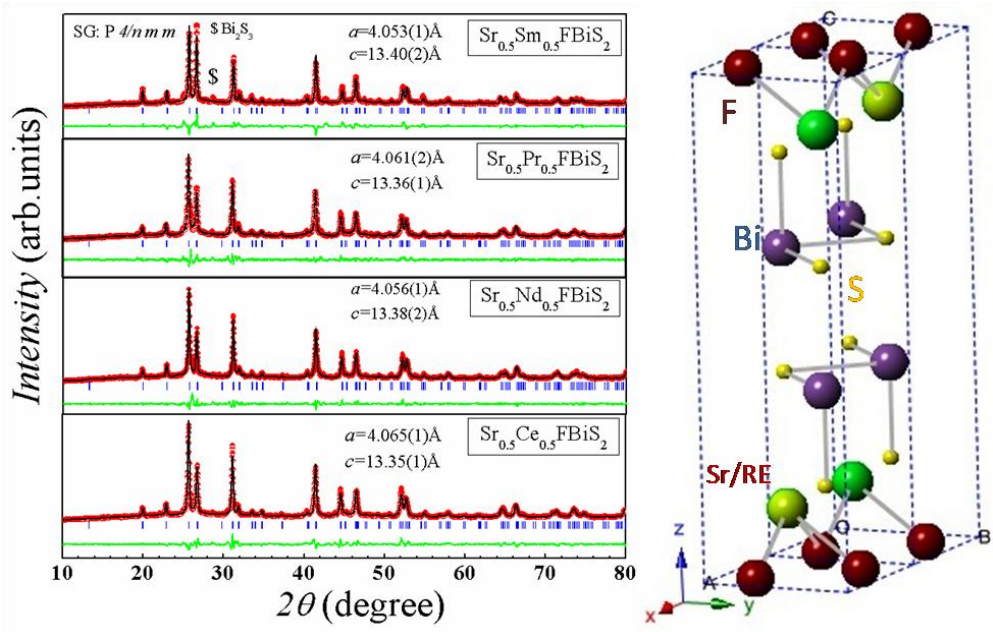

**Fig. 2**

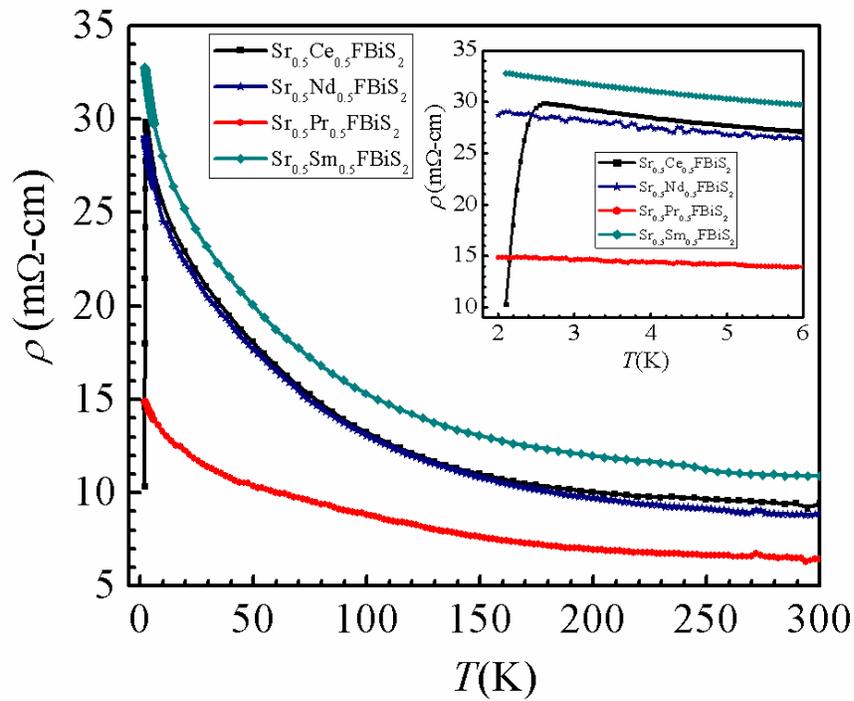

**Fig. 3**

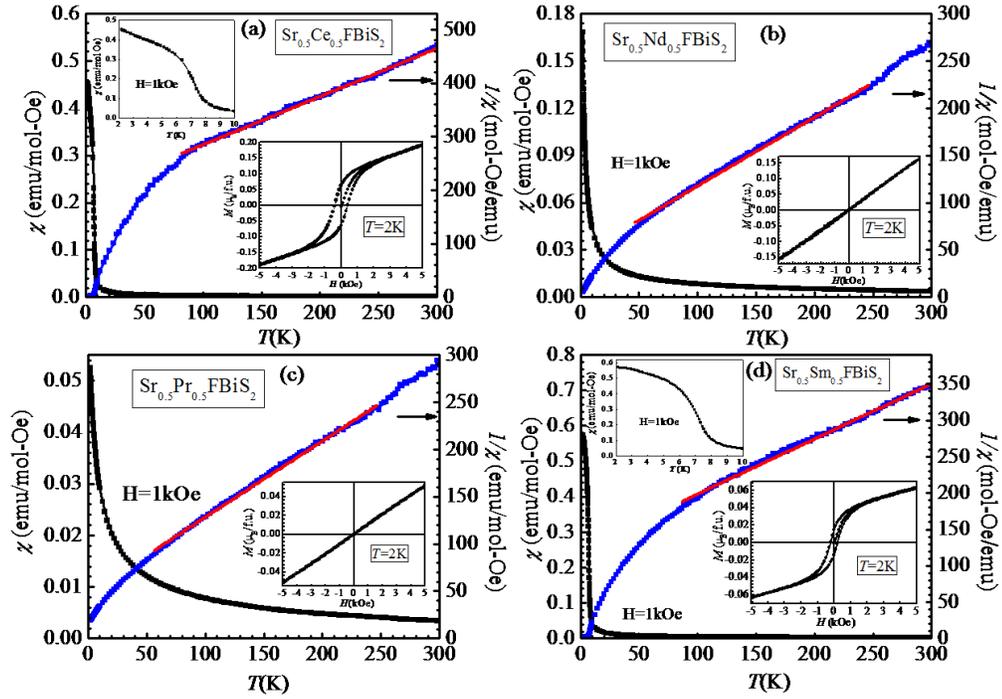

**Fig. 4**

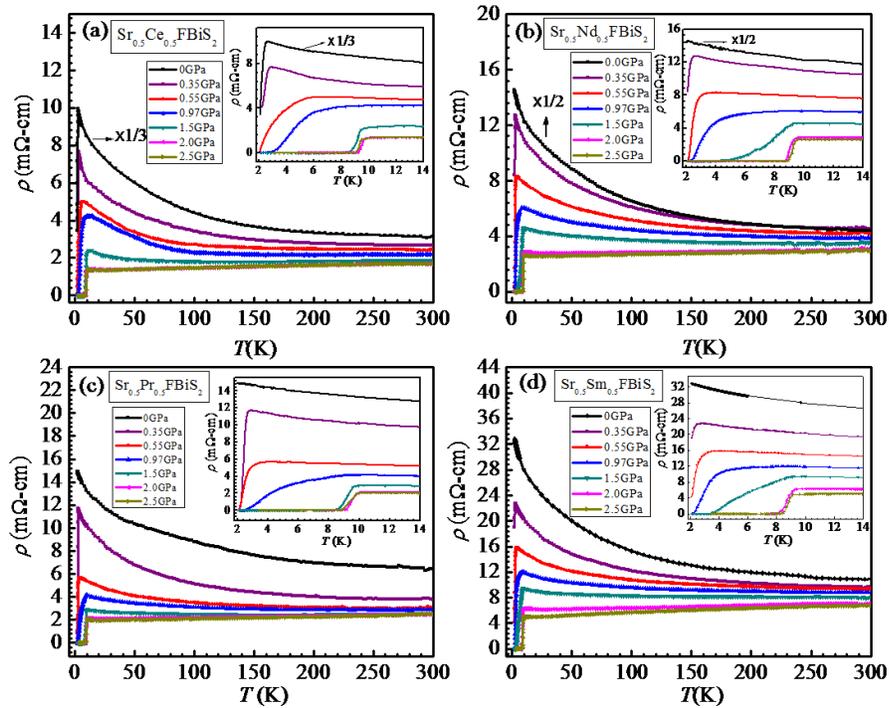

**Fig. 5**

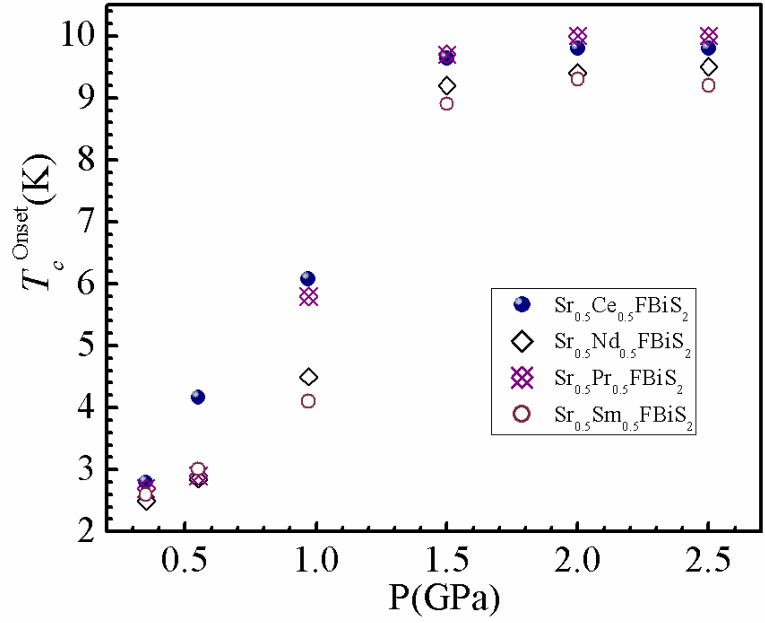

**Fig. 6**

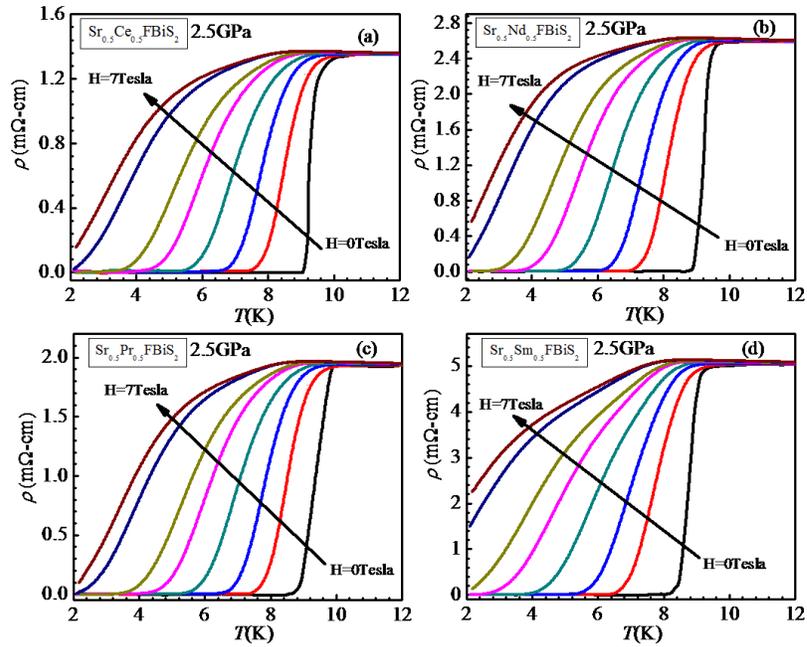

Fig. 7

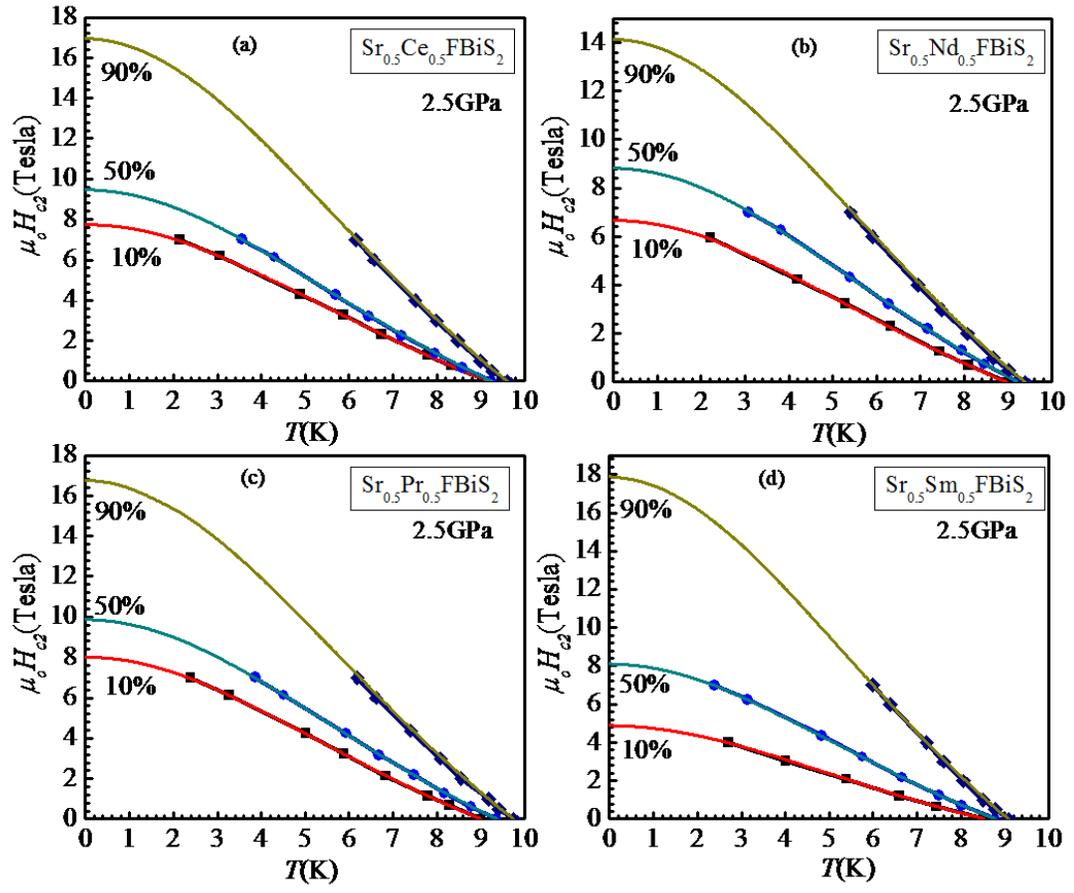